\documentclass[aps,prb,twocolumn,superscriptaddress]{revtex4-2}
\usepackage{bm}
\usepackage{ae}
\usepackage[T1]{fontenc}
\usepackage[ansinew]{inputenc}
\usepackage{amsmath}
\usepackage{amssymb}
\usepackage{graphicx}
\usepackage{color}
\usepackage[colorlinks]{hyperref}
\usepackage{epstopdf}
\input{epsf}

\begin{document}

\title{Transduction of quantum information from charge qubit to nanomechanical cat-state}

\author{D. Radi\'{c}*}
\affiliation{Department of Physics, Faculty of Science, University of Zagreb, Bijeni\v{c}ka 32, Zagreb 10000, Croatia}

\author{L.Y.Gorelik}
\affiliation{Department of Physics, Chalmers University of Technology, SE-412 96 G{\"o}teborg, Sweden}

\author{S.I.Kulinich}
\affiliation{B. Verkin Institute for Low Temperature Physics and Engineering of the National Academy of Sciences of Ukraine, 47 Prospekt Nauky, Kharkiv 61103, Ukraine}

\author{R. I. Shekhter}
\affiliation{Department of Physics, University of Gothenburg, SE-412 96 G\"oteborg, Sweden}


\begin{abstract}

We suggest a nanoelectromechanical setup and corresponding time-protocol of its manipulation by which we transduce quantum information from charge qubit to nanomechanical cat-state. The setup is based on the AC Josephson effect between bulk superconductors and mechanically vibrating mesoscopic superconducting island in the regime of the Cooper pair box. Starting with a pure state with quantum information initially encoded into superposition of the Cooper pair box states, applying a specially tailored time-protocol upon bias voltage and gate electrodes, we obtain a new pure state with information finally encoded into superposition of nanomechanical coherent states constituting the cat-state. This performance is achieved using quantum entanglement between electrical and mechanical states. Nanomechanical cat-states serve as a "storage" of quantum information, motivated by significantly longer decoherence time with respect to the charge qubit states, from which the information can be transdued back to the charge qubit applying the reverse time-protocol. 
\end{abstract}

\maketitle

\section*{Introduction}

The implementation of interfaces that store quantum information (QI) is in the focus of modern research in the field of quantum communication. The basic "container" of the QI is a qubit \cite{Schumacher}, the system with two quantum states where the QI is stored in their superposition \cite{Nielsen}.
The implementation of such 2-level system is a subject of a huge area of research, covering number of fields in physics such as optics, atomic physics or solid-state physics, as well as their combinations \cite{QuantumProcessing,Girvin,Devoret,Mirhosseini,Leibfried}, where one of the key questions is ability of a qubit to preserve encoded information, i.e. its resilience to different deterioration processes due to the interaction with environment. In our research we are focused to the nanoelectromechanical (NEM) implementation of superconducting qubit and transduction of the QI to nanomechanics, for which the coherent interplay of qubit states and nanomechanical excitations is essential \cite{Schneider,Hann,Chu}.

One successful implementation of such transduction was recently demonstrated experimentally by coupling of a superconducting qubit circuit to a mechanical surface resonator \cite{Cleland1,Cleland2}. There, the individual phonons can be controlled and detected by a superconducting qubit circuit, enabling a coherent generation and measurement of a non-classical superposition of the zero- and one-phonon itinerant Fock states. This control allows one to accomplish phonon-mediated quantum state transfer and establish remote qubit entanglement.

In our previous publications \cite{npjDR,Bahrova} we suggested a NEM setup in which we coupled charge degrees of freedom of the superconducting mesoscopic grain, in the regime of the Cooper pair box (CPB qubit) \cite{Bouchiat,Nakamura,Robert,Lehnert}, with its nanomechanical vibrations between the superconducting leads biased by a constant voltage. Such a setup, based on the AC Josephson effect between the superconducting leads and the CPB qubit, resulted in formation of cat-states consisting of nanomechanical coherent states entangled with the states of the CPB if Josephson frequency and mechanical frequency are in resonance. In such a setup, the QI is stored in entangled qubit and mechanical states, being susceptible to its deterioration due to rather short-living coherence of the charge qubit \cite{Nakamura}. Entanglement itself is a powerful resource as for this particular type of functionality, but also in the field of quantum communication and computation in general \cite{Horodecki}.  

In this paper we are focused to transduction of quantum information between charge and mechanical subsystems. Based on the previous papers \cite{npjDR,Bahrova}, we suggest a modified setup and corresponding time-protocol of manipulating the control parameters in order to achieve this goal. In particular, we start with a prepared initial state $(\alpha \mid \uparrow \rangle + \beta \mid \downarrow \rangle ) \otimes \vert 0 \rangle$, $\vert \alpha\vert^2+\vert\beta\vert^2 =1$, with the QI encoded in superposition of the CPB states $\mid \uparrow \rangle$ and $\mid \downarrow \rangle$ in terms of complex coefficients $\alpha$ and $\beta$ (i.e. in their relative phase), where $\vert 0 \rangle$ is nanomechanical vacuum state.
Utilizing our time-protocol, we finally want to obtain the pure state $\mid \downarrow \rangle \otimes (\alpha \vert Z \rangle + \beta \vert -Z \rangle )$, where QI is encoded into special superposition of mechanical coherent states $\vert \pm Z \rangle$ known as a cat-state.

The paper is organized in the following way: in Section I we present the setup and model Hamiltonian of the system; in Section II we present the time-protocol, calculate the time-evolution operator and wave function; in Section III we calculate the corresponding Wigner function and entropy of entanglement; section Conclusions contains the concluding remarks and discussion; in section Appendix some mathematical procedures are derived.

\section{The Model}

\begin{figure}
\centerline{\includegraphics[width=\columnwidth]{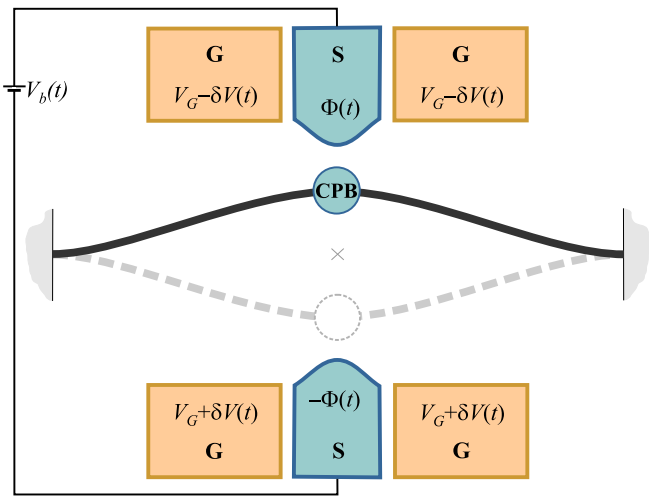}}
\caption{Schematic illustration of the NEM setup. The superconducting mesoscopic quantum dot (CPB) is attached to a suspended nanowire (e.g. a carbon nanotube), capable of performing the in-plane harmonic vibrations between two superconducting electrodes (S) symmetrically biased by a constant voltage $V_b$ used to create a superconducting phase $\Phi(t)$. Electrostatic gate electrodes (G) usage is twofold: $V_G$ is used to bring the states of the quantum dot with zero and one excess Cooper pair  into the CPB regime, while $\delta V(t)$ is used in the time-protocol to provide a finite approximately homogeneous electric field across the central region of the junction (cross) on demand, while changing no potential (due to symmetric configuration).}
\label{FigSchematic}
\end{figure}

Schematic picture of the proposed nanomechanical setup is shown in Fig.\ref{FigSchematic}. 
%
%
%
%
By the particular choice of potential $V_G$ on gate electrodes (G), the mesoscopic superconducting island (quantum dot) is set into regime of the Cooper pair box, i.e. the effective two-level system of degenerate states with zero and one excess Cooper pair, denoted by $\mid \downarrow \rangle$ and $\mid \uparrow \rangle$ respectively. Those we call "the charge states" or "the CPB qubit states". The junction operates in regime of the AC Josephson effect, i.e. the superconducting electrodes (S) are biased by a constant symmetric bias voltage $V_b$, providing there a superconducting phase $\pm \Phi(t)$, where $\Phi(t)=\Omega t + \Phi_0$, $\Omega = 2eV_b/\hbar$ is the Josephson frequency and $\Phi_0$ is some constant initial phase. The Cooper pairs tunnel between the S-electrodes and the CPB which is attached to a nanowire, performing mechanical vibrations at frequency $\omega$, thus making the tunneling strongly position-dependent. Symmetric configuration of the G-electrodes provides a possibility, by applying a constant additional voltage $\delta V$ with a proper sign, to create an approximately homogeneous electric field $\mathcal{E}$ along the central part of the junction where the CPB moves, without changing the potential on the CPB (i.e. preserving the degeneracy of states).

Expanding the Josephson coupling in terms of the small parameter $\varepsilon \equiv x_0 / x_{\text{tun}} \ll 1$, where $x_0=\sqrt{h/m\omega}$ is the amplitude of zero-mode oscillations ($m$ is mass of the oscillator) and $x_{\text{tun}}$ is the tunneling length, we adopt the time-dependent model Hamiltinian
%
\begin{eqnarray}\label{Hamiltonian}
\hat H(t)&=&E_J\sigma_1\cos \Phi(t)+\hbar\omega b^\dag b
+\varepsilon E_J(b+b^\dag) \sigma_2 \sin \Phi (t) \nonumber\\
&+& E(t)(b+b^\dag)\sigma_3.
\end{eqnarray}
%
Here $E_J$ is the Josephson energy, $E(t)=2|e|\mathcal{E}(t)x_0$ is the (eletrostatic) energy related to electric field, $\sigma_{1,2,3}$ are the Pauli matrices operating in the CPB qubit $2 \times 2$ space, while $b^{\dagger}$ is the phonon creation operator. The Hamiltonian (\ref{Hamiltonian}) is written in so-called charge representation where the qubit states are $\mid \downarrow \rangle = (0,1)^{\text{T}}$ and $\mid \uparrow \rangle=(1,0)^{\text{T}}$ .

\section{The time-protocol}

The control mechanisms upon the proposed setup at our hands are in terms of voltages $V_b$, $V_G$ and $\delta V$ and the time moment we switch them on/off. In that sense they are functions of time. As said before, $V_G$ is set all the time at a particular value to keep the superconducting island in the CPB regime. Our aim is to, starting from the initial state described by wave function $\vert \mathbf{\Psi}(t=0) \rangle=\mathbf{e}_{\text{in}} \otimes \vert 0 \rangle$, where $\mathbf{e}_{\text{in}}=\alpha \mid \uparrow \rangle + \beta \mid \downarrow \rangle$, construct the time-protocol of operation with $V_b(t)$ and $\delta V(t)$ that will yield the final state with QI encoded in the mechanical cat-state as described in the Introduction. In that respect we define two characteristic moments in time, $t_0$ and $t_s$. The moment $t_0$ is defined through equality $\Omega t_0 + \Phi_0 = \pi/2$, i.e. it is determined by parameters $\Omega$ and $\Phi_0$. The moment $t_s$ is defined as $t_s = t_0 + NT$, where $T=2\pi/\Omega$ is period and $N$ is an integer. Operation upon voltages is the following: at $t=0$ we switch on the constant bias voltage $V_b$ while keeping $\delta V=0$, keeping those values along the time interval $t\in(0,t_s)$. Therefore, during that time interval the phase $\Phi(t)$ is increasing linear in time, while $E(t)=0$. At the moment $t=t_s$, when the phase attains a value $\Phi(t_s)=2\pi N+\pi/2$, we switch off the bias voltage and switch on the constant $\delta V$, consequently providing a constant electric field $\mathcal{E}$. As the bias voltage is switched off, the phase remains constant, i.e. $\Phi(t>t_s)=2\pi N+\pi/2=const$, while the position-dependent evolution of qubit states takes place with our aim to bring them to coincidence. Position-dependence of evolution operator is important because we use the fact that coherent states evolve in the regions of the phase space with opposite sign of coordinate, so the unitary rotation of qubit states, associated with each coherent state, will be "in opposite direction".  

Based on our analysis \cite{npjDR} we restrict consideration to the resonant case $\Omega = \omega$ embracing a coherent dynamics of Josephson tunneling and mechanical oscillator.

In order to achieve the sought functionality we should construct the time-evolution operator based on Hamiltonian (\ref{Hamiltonian}) and defined time-protocol of engaging terms within it. The first step is to determine it at the time-interval $(0,t_s)$. 
The evolution operator $\hat U_{\text{I}}(t,t')$, for times $t,t'\in (0,t_s)$ was determined in \cite{npjDR}. Using the interaction representation, we look for the evolution operator in the form
%
\begin{eqnarray}\label{U_I-1}
\hat U_{\text{I}}(t,t')=\hat U_0(t)\hat u(t,t')\hat U_0^\dag (t'),
\end{eqnarray}
%
in which 
%
\begin{eqnarray}\label{U_0}
\hat U_0(t)= \exp\left[-\imath \omega t b^\dag b-\imath
\kappa\sigma_1\sin\Phi(t)\right],
\end{eqnarray}
%
where $\kappa \equiv E_J/\hbar\omega$ and $\hat u(t,t)=\hat I$. Then, the equation for the evolution operator $\hat u(t,t')$ attains the form
%
\begin{eqnarray}\label{u_equation}
\imath \hbar\frac{\partial \hat u(t,t')}{\partial
t}=\hat H_{\text{eff}}(t) \hat u(t,t'),
\end{eqnarray}
%
where
%
\begin{eqnarray}\label{H_eff}
\hat H_{\text{eff}}(t)=\frac{\varepsilon E_J}{4}\left[\imath
b^\dag e^{-\imath \Phi_0} +\imath b e^{-2\imath \omega t -\imath
\Phi_0} +\text{h.c.}\right] \times \nonumber\\
\left[(\sigma_2+\imath\sigma_3)
e^{2\imath \kappa \sin \Phi(t)}+\text{h.c.}\right].
\end{eqnarray}
%
When the resonance condition $\omega=\Omega$ is fulfilled, the
effective Hamiltonian $\hat H_{\text{eff}}(t)$ is a periodic function with the same period $T$ as the original Hamiltonian Eq.(\ref{Hamiltonian}). As a consequence, the evolution operator $\hat u(t,t')$ has a property of discrete translational invariance, i.e. $\hat u(t-T,t'-T)=\hat u(t,t')$.
Using that property, we can write the evolution operator in the form
%
\begin{eqnarray}\label{u_T-product}
\hat u(t_s,0)&=& \hat u(t_s,t_s-T) \hat u(t_s-T,t_s-2T)\cdots \nonumber\\
&& \hspace{2.5cm} \cdots \hat u(t_0+T,t_0) \hat u(t_0,0) \nonumber\\
&=&\hat u^N(t_0+T,t_0) \hat u(t_0,0).
\end{eqnarray}
%
Due to the smallness of parameter $\varepsilon$, one can put
$\hat u(t_0,0) \simeq \hat I$. Within framework of the rotating wave approximation (RWA), the Eq. (\ref{u_equation}) yields the operator $\hat u(t_0+T,t_0)$ in the form
%
\begin{equation}\label{u_operator}
\hat u(t_0+T,t_0)=\hat I+\lambda \sigma_2\left(b^\dag
e^{-\imath\Phi_0}-be^{\imath\Phi_0}\right)+{\cal{O}}(\varepsilon^2),
\end{equation}
%
where $\lambda=\pi\varepsilon\kappa (J_0(2\kappa)-J_2(2\kappa))$, $J_i(2\kappa)$ is the Bessel function of the first kind.

Using Eqs.(\ref{U_I-1}), (\ref{u_T-product}) and (\ref{u_operator}),
one finally obtains the time-evolution operator
%
\begin{eqnarray}\label{U_I-final}
&& \hspace{-0.5cm} \hat U_{\text{I}}(t_s,0) = \nonumber\\
&=&e^{-\imath\kappa\sigma_1}e^{-\imath\omega t_s b^\dag b}\left[1+\lambda \sigma_2 \left(b^\dag
e^{-\imath\Phi_0}-be^{\imath\Phi_0}\right)\right]^N \times \nonumber\\
&& \hspace{5.5cm} e^{\imath \kappa\sigma_1\sin\Phi_0}\nonumber\\
&=&e^{-\imath\kappa\sigma_1}\left[1+\lambda \sigma_2 \left(b^\dag
e^{-\imath(\omega t_s+\Phi_0)}-be^{\imath(\omega
t_s+\Phi_0)}\right)\right]^N \times \nonumber\\
&& \hspace{4.5cm} e^{\imath\kappa\sigma_1\sin\Phi_0} 
e^{-\imath\omega t_s b^\dag b} \nonumber\\
&\simeq & e^{-\imath\kappa\sigma_1}e^{-\imath N
\lambda\sigma_2(b+b^\dag)}e^{\imath\kappa\sigma_1\sin\Phi_0}
e^{-\imath\omega t_s b^\dag b}.
\end{eqnarray}
%
derived under condition $\varepsilon^2 N \ll 1$.\\

The second step is to determine a time-evolution operator in the interval $t>t_s$ in which the Hamiltonian (\ref{Hamiltonian}) is time-independent. There, one can neglect the small (third) term proportional to $\varepsilon \ll 1$ accounting for Josephson coupling of qubit to mechanics. The evolution operator is
%
\begin{eqnarray}\label{U_II-1}
\hat U_{\text{II}}(t,t_s)&=&\exp\left[-\imath\tau\hat
H/\hbar\right] \nonumber\\
&\simeq & \exp\left\{-\imath \tau\left[\omega b^\dag
b+\sigma_3(E/\hbar)(b+b^\dag)\right]\right\},
\end{eqnarray}
%
where $\tau \equiv t-t_s$. Using results presented in Appendix,
one can rewrite expression (\ref{U_II-1}) in the form
%
\begin{eqnarray}\label{U_II-final}
\hat U_{\text{II}}(t,t_s)&=& f(\tau)\exp\left[-\xi\left(1-e^{-\imath
\omega\tau}\right)\sigma_3 b^\dag\right] e^{-\imath\omega\tau
b^\dag b} \times \nonumber\\
&& \hspace{2cm} \exp\left[ -\xi\left(1-e^{-\imath \omega\tau}\right)
\sigma_3 b\right],
\end{eqnarray}
%
where $\xi=E/\hbar\omega$ and
%
\begin{equation}\label{f-function}
f(\tau)=\exp\left[-\xi^2 \right(1-\imath\omega\tau-
e^{-\imath\omega\tau} \left)\right].
\end{equation}
%

Expressions (\ref{U_I-final}) and (\ref{U_II-final}), defining the evolution operator for the sought time-protocol, determine the wave function $\vert \mathbf{\Psi}(t)\rangle$ at any moment of time $t>t_s$. The result is
%
\begin{equation}\label{Psi_entangled}
\vert \mathbf \Psi(t)\rangle= e^{\imath \varphi_0}\left(\mathbf
e_3^+ \otimes\vert C_1\rangle +\mathbf e_3^-\otimes\vert
C_2\rangle\right),
\end{equation}
%
where vector $\mathbf e_j^\pm$, $j=1,2,3$, is an eigenvector of Pauli
matrix $\sigma_j$ with eigenvalues $\pm 1$,
%
\begin{eqnarray}\label{C_coeffs}
\vert C_1(\tau)\rangle &=& A_+\sin\vartheta
e^{\imath\varphi_1}\vert q_+\rangle+A_-\cos\vartheta
e^{-\imath\varphi_1}\vert
-q_-\rangle, \nonumber\\
\vert C_2(\tau)\rangle &=& A_+\cos\vartheta
e^{-\imath\varphi_1}\vert q_-\rangle+A_-\sin\vartheta
e^{\imath\varphi_1}\vert -q_+\rangle, \nonumber\\
\end{eqnarray}
%
in which
%
\begin{eqnarray}\label{q_PM}
q_{\pm}(\tau) &=& -\imath\lambda N
e^{-\imath\omega\tau}\mp\xi\left(1-e^{-\imath\omega\tau}
\right),
\end{eqnarray}
%
and $\vartheta \equiv \kappa+\pi/4$. Here $\vert \pm q_\pm \rangle$ are mechanical coherent states defined as $\vert Z \rangle \equiv \exp\left[Zb^\dag-Z^\ast b\right]\vert 0\rangle$.
Functions $\varphi_j$, $j=0,1$ are defined by
%
\begin{eqnarray}\label{fi_functions}
&&\varphi_0(\tau)=\xi^2\left(\omega\tau-\sin\omega\tau\right)+\pi/4, \nonumber\\
&&\varphi_1(\tau)=\lambda \xi N(1-\cos\omega\tau)-\pi/4,
\end{eqnarray}
%
while constants $A_\pm$ are determined by the initial condition, i.e.
%
\begin{equation}\label{A_coeffs}
A_\pm=\mathbf e_2^\pm e^{\imath\kappa\sin \Phi_0\sigma_1}\mathbf{e}_{\text{in}},
\end{equation}
%
fulfilling the unitarity condition $\vert A_+\vert ^2+\vert
A_-\vert^2=1$.
The wave function (\ref{Psi_entangled}) presents an entangled state of qubit and mechanical subsystem.

We consider a special case fulfilling conditions
%
\begin{equation}\label{ValidityConditions}
\xi\ll 1, \hspace{0.3cm} \lambda N\gg 1, \hspace{0.3cm} \xi\lambda N\sim 1,
\end{equation}
%
under which we can make approximation in Eq. (\ref{q_PM}), i.e.
%
\begin{equation}\label{q_0}
q_+\simeq q_-\simeq -\imath \lambda N e^{-\imath\omega\tau} \equiv q_0.
\end{equation}
%

\begin{figure}
\centerline{\includegraphics[width=\columnwidth]{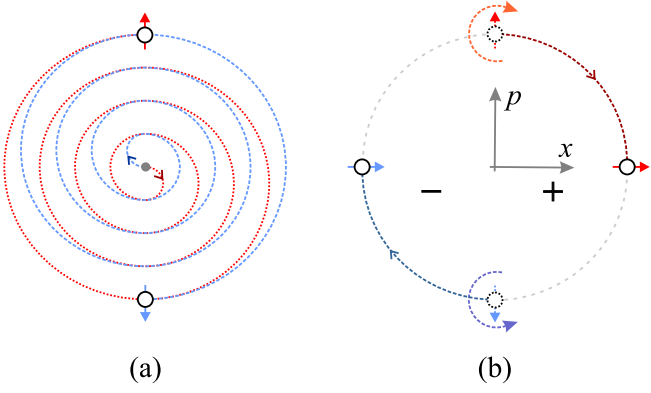}}
\caption{Schematic illustration of the QI transduction protocol in the $(x,p)$ phase space. (a) During the time interval $t\in(0,t_s)$ the entangled state Eq. (\ref{Psi_entangled}) of qubit and mechanical coherent states is built applying the coherent AC Josephson effect-based dynamics coupled to mechanical oscillator while keeping electric field equal to zero. (b) During the time interval $t\in(t_s,t_s+\tau_\ast)$ the pure state Eq. (\ref{Psi_final}) is built by switching off the bias voltage (Josephson dynamics) and applying the finite electric field $\mathcal{E}$ performing "rotation" of qubit states in opposite directions, due to position-dependent $(+/-)$ interaction, until they coincide.}
\label{FigTransduction}
\end{figure}

If we set the value of ratio $\kappa$ to be equal to $2\pi M$, where $M$ is an integer, then we can define a special moment of time $\tau=\tau_\ast$, where $\tau_\ast$ is determined by the relation $\xi\lambda N(1-\cos\omega\tau_\ast)=\pi/4$. At the time moment $\tau=\tau_\ast$ the wave function (\ref{Psi_entangled}) takes the form
%
\begin{equation}\label{Psi_final}
\vert \mathbf \Psi(\tau=\tau_\ast)\rangle\simeq e^{\imath \varphi_0} \mathbf
e_1^+\otimes \left(A_+ \vert q_0\rangle+A_-\vert
-q_0\rangle\right),
\end{equation}
%
when we can switch off the electric field (i.e. set $\delta V=0$) preventing further evolution from this form. 
We see that state of the system has changed to the "opposite form" - the electronic subsystem is now in a certain state $\mathbf{e}_1^+$, while the mechanical subsystem transformed into a superposition of two coherent states, with coefficients $A_+$ and $A_-$, constituting the cat-state. It exactly fulfills our goal  to get the pure state with the QI transduced from the electronic subsystem to the mechanical one. The process is schematically depicted in Fig. \ref{FigTransduction}.
%
%
%
%

\section{The Wigner function and entropy of  entanglement}

Using Eq. (\ref{Psi_entangled}), we construct a reduced density matrix of the mechanical subsystem $\hat \rho_m(t)=\text{Tr}_q \hat\rho (t)$, where $\hat \rho=\vert \mathbf \Psi(t)\rangle\langle\mathbf \Psi(t)\vert$ is the complete density matrix of the system, which takes the form
%
\begin{equation}\label{rho_m}
\hat \rho_m(t)=\vert C_1\rangle\langle C_1\vert+\vert
C_2\rangle\langle C_2\vert
\end{equation}
%
in terms of coefficients defined by Eqs. (\ref{C_coeffs}) and (\ref{q_0}).
To describe the evolution of mechanical subsystem it is convenient
to use the Wigner function representation for the density matrix $\hat \rho_m(t)$
%
\begin{equation}\label{Wigner_0}
W(x,p,t)=\frac{1}{2\pi}\int e^{-\imath p \zeta} \langle
x+\tfrac{1}{2}\zeta\vert\hat\rho_m(t)\vert x-\tfrac{1}{2}\zeta \rangle d\zeta.
\end{equation}
%
The direct calculations show that it is convenient to present the Wigner function as a sum of two terms, $W=W_1+W_2$. The first term, $W_1(x,p,\tau)$, describes the evolution of the four coherent states
%
\begin{eqnarray}\label{Wigner_1}
W_1(x,p,\tau)&=&\tfrac{1}{\pi} \vert A_+\vert^2 \left\{\sin^2\vartheta\,
e^{-[x-x_1(\tau)]^2-[p-p_1(\tau)]^2} \right. \nonumber\\
&&\left. +\cos^2\vartheta \, e^{-[x-x_2(\tau)]^2-[p-p_2(\tau)]^2}\right\} \nonumber\\
&&+\tfrac{1}{\pi}\vert A_-\vert^2 \left\{\sin^2\vartheta \,
e^{-[x+x_1(\tau)]^2-[p+p_1(\tau)]^2} \right. \nonumber\\
&&\left. +\cos^2\vartheta \, e^{-[x+x_2(\tau)]^2-[p+p_2(\tau)]^2}\right\},
\end{eqnarray}
%
where
%
\begin{eqnarray}\label{CAT-state_positions}
x_{1,2}(\tau) &=& -\sqrt 2\left[\lambda
N\sin\omega\tau\pm\xi(1-\cos\omega\tau)\right], \nonumber\\
p_{1,2}(\tau) &=& -\sqrt 2\left(\lambda N\cos\omega
\tau\pm\xi\sin\omega\tau\right)
\end{eqnarray}
%
determine their positions in the phase space.

The second term, $W_2(x,p,\tau)$, describes the interference
pattern arising under the such dynamics
%
\begin{eqnarray}\label{Wigner_2}
W_2(x,p,\tau)&=&\imath\tfrac{1}{2\pi} A_+A_-^\ast\, \sin
2\vartheta \, e^{\imath\Theta(x,p,\tau)} \times \nonumber\\
&&\left\{e^{-[x-x_0(\tau)]^2-
[p-p_0(\tau)]^2 -\imath \Phi_1(\tau)} \right. \nonumber\\
&&\left. -e^{-[x+x_0(\tau)]^2
-[p+p_0(\tau)]^2+\imath \Phi_1(\tau)}\right\}+\text{c.c.},\nonumber\\
\end{eqnarray}
%
where
%
\begin{eqnarray}\label{interference_positions}
x_0(\tau)&=&\frac{x_2(\tau)-x_1(\tau)}{2}=\sqrt 2
\xi(1-\cos\omega\tau), \nonumber\\
p_0(\tau)&=&\frac{p_2(\tau)-p_1(\tau)}{2}=\sqrt 2 \xi
\sin\omega\tau,
\end{eqnarray}
%
and
%
\begin{eqnarray}\label{interference_phases}
\Theta(x,p,\tau)&=&2\sqrt 2 \lambda N
(p\sin\omega\tau-x\cos\omega\tau), \nonumber\\
 \Phi_1(\tau)&=&4\xi\lambda
N(1-\cos\omega\tau).
\end{eqnarray}
%

Using Eq.(\ref{Psi_entangled}), we now construct a reduced density matrix of the qubit subsystem  $\hat \rho_q(t)=\text{Tr}_m \hat\rho (t)$ which takes the form
%
\begin{equation}\label{rho_q}
\hat\rho_q=\frac{1+\hat A}{2},
\end{equation}
%
where
%
\begin{eqnarray}\label{A_matrix}
\hat A(\tau)&=&[\langle C_1,C_1\rangle-\langle
C_2,C_2\rangle]\sigma_3 \nonumber\\
&&\hspace{1.2cm}+2\langle C_1,C_2\rangle\sigma_-+2\langle C_2,C_1\rangle\sigma_+
\end{eqnarray}
%
and $\sigma_\pm = (\sigma_1\pm \imath \sigma_2)/2$.
The entropy of entanglement $S_{\text{en}}=\text{Tr}(\hat \rho_q \log \hat \rho_q)$ then equals to
%
\begin{equation}\label{Entropy}
S_{\text{en}}(\tau)=\ln 2-\frac{(1+\epsilon)\ln(1+\epsilon)+
(1-\epsilon)\ln(1-\epsilon)}{2},
\end{equation}
%
where
%
\begin{equation}\label{parameter_epsilon_def}
\epsilon^2(\tau)=[\langle C_1,C_1\rangle-\langle
C_2,C_2\rangle]^2+4\vert\langle C_1,C_2\rangle\vert^2.
\end{equation}
%
Within the considered regime, defined by Eq (\ref{ValidityConditions}), the expression for the parameter $\epsilon$ takes an approximate form
%
\begin{eqnarray}\label{parameter_epsilon_approx}
\epsilon^2(\tau) \simeq 1 &-& 4\vert A_1 A_2\vert^2 \, (1 \nonumber\\
&-& \sin^2 2\vartheta\sin^2[4\xi\lambda N(1-\cos \omega\tau)]).
\end{eqnarray}
%

\begin{figure}
\centerline{\includegraphics[width=\columnwidth]{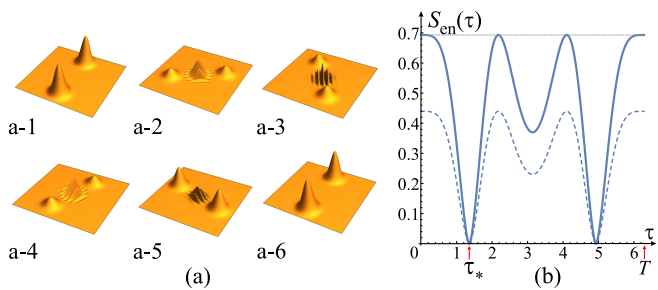}}
\caption{(a) The Wigner function $W(x,p,\tau)=W_1(x,p,\tau)+W_2(x,p,\tau)$, defined by Eqs. (\ref{Wigner_1}) and (\ref{Wigner_2}) under conditions (\ref{ValidityConditions}), in the $(x,p)-$phase space depending on time $\tau\equiv=t-t_s=0$, $1$, $2.5$, $3.5$, $4.5$, $2\pi$ in unites of $\omega^{-1}$ shown in figs. a-1 to a-6 respectfully. (b)The entropy of entanglement Eq. (\ref{Entropy}) as a function of time $\tau$ (in units $\omega^{-1}$, $T=2\pi/\omega$ is a period of oscillations) under conditions (\ref{ValidityConditions}). The dashed horizontal line a maximal possible value of entropy $\ln(2)$. The moment $\tau_\ast$ is the first moment at which the pure state is achieved, thus $S_{\text{en}}(\tau_\ast)=0$. The parameters used were $\vert A_+ \vert=\vert A_- \vert=1/\sqrt{2}$, $\xi\lambda N \approx 1$, $\theta =2\pi M+\pi/4$, $M$ is an integer. The dashed curve in (b) is calculated for $\vert A_+ \vert=0.4$,  $\vert A_- \vert=0.9165$.}
\label{FigResults}
\end{figure}

Therefore, in the time interval $t>t_s$ one obtains an oscillatory dependence of the entropy Eq. (\ref{Entropy}) on time with the period $T=2~pi/\omega$. It reflects a rotation between pure and entangled states. The maxima in the entaglement entropy correspond to maximally entangled states for which the interference pattern in the Wigner function, Eq. (\ref{Wigner_2}), vanishes.
Notice also that, in the case $A_+=0$ or $A_-=0$, the entropy of entanglement vanishes reflecting the fact that we have a pure state under consideration. The Wigner function interference pattern vanishes as well since there is no quantum superposition present. Results are shown in Fig. \ref{FigResults}.
%
%

\section{Conclusions}

\begin{figure}
\centerline{\includegraphics[width=0.85\columnwidth]{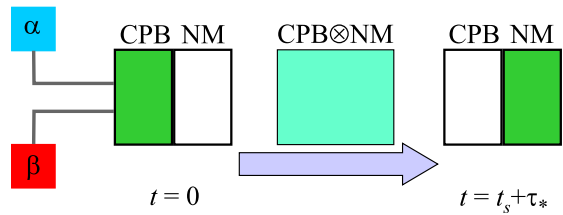}}
\caption{Schematic illustration of transduction process. The QI ($\alpha$, $\beta$) is encoded into the charge qubit (CPB) at $t=0$ while nanomechanical subsystem (NM) is in the vacuum state. During execution of the protocol it is stored in the entangled states of charge qubit and nanomechanics (CPB$\otimes$NM). At $t=t_s+\tau_\ast$, the QI is transduced into the NM subsystem.}
\label{FigProtocol}
\end{figure}

We propose a nanoelectromechanical setup, based on the AC Josephson effect between superconducting electrodes and superconducting mesoscopic island, in the regime of Cooper pair box (charge qubit), performing the nanomechanical oscillations between them. We control the junction by the bias voltage and specially arranged gate voltages (see Fig. \ref{FigSchematic}) in the resonant regime of the Josephson and mechanical frequency. The initial state of the system is direct product of the mechanical vacuum state and prepared quantum superposition of qubit states into which the quantum information is encoded. For such setup we propose a time-protocol to operate mentioned voltages by which we time-evolve the wave function. In the first time interval, directed by the bias voltage, we build the entangled state of the charge qubit and mechanical coherent states. In the second interval, directed by the gate voltage that generates finite electric field acting on qubit by the position-dependent coupling, we time-evolve the wave function into a pure state which is direct product of certain qubit state and mechanical cat-state with initial quantum information transduced into superposition of nanomechanical coherent states. Applying a "reverse" time-protocol, we may perform the "reverse transduction", i.e. return the quantum information into superposition of qubit states again. Schematically depicted protocol of transduction is shown in Fig. \ref{FigProtocol}.
%
%

Cat-states have been widely used in the field of quantum optics so far \cite{Hacker}. From the point of view of quantum information, their property to provide protocols for quantum error correction \cite{Girvin} is very intriguing. We suggest an implementation in the field of solid-states hybrid systems, particularly in the field of nanomechanics where we utilize a cat-state as a "nanomechanical qubit". Motivation to transduce the QI into nanomechanics is obvious - it is a platform capable of providing huge quality factors, growing every day \cite{Laird,Tao,Bereyhi}. Additionally, the nanomechanical coherent states, which make the cat-state, are the robust multiphonon states, rather resilient to decoherence processes such as phonon loss for example. The charge qubit, that we couple to the nanomechanical system, is, on the other hand, very convenient for the external control mechanisms based on interaction with charge, or quantum-mechanical tunneling. Unfortunately, exactly that sensitivity makes it very vulnerable with respect to the external "Coulomb noise" yielding the way shorter decohorence time  comparing to mechanics \cite{Nakamura}. That is where the idea to utilize charge qubit as a control terminal for the QI encoding, and then transduce it into "mechanical storage" and back by demand, is founded. It constitutes the first step along our research direction with a goal to transfer the QI in space between different qubits utilizing the nanomechanics.

\section*{Appendix}

Let us define $\zeta \equiv -\imath\tau/\hbar$ and consider the expression
%
\begin{equation}\label{100}
\exp \zeta\left(\hat A+\hat B+\hat C\right)= \delta(\zeta)e^
{\beta(\zeta)\hat B}e^{\alpha(\zeta)\hat A}e^{\gamma(\zeta)\hat
C},
\end{equation}
%
where $\hat A =\hbar\omega b^\dag b$, $\hat B=  E \sigma_3 b^\dag$,
$\hat C=  E \sigma_3 b$, so $\left[\hat A, \hat
B\right]=\hbar\omega\hat B$, $\left[\hat A, \hat C\right]=-\hbar\omega\hat C$, $\left[\hat B, \hat C\right]=-E^2$, and $\alpha(\zeta)$, $\beta(\zeta)$, $\gamma(\zeta)$, $\delta(\zeta)$ are
the sought functions. Performing the differentiation on Eq.(\ref{100}) with respect to $\zeta$ and multiplying the resulting expression from the right by the operator
%
\begin{equation}\label{101}
\exp \left[-\zeta\left(\hat A+\hat B+\hat
C\right)\right]=\delta^{-1}(\zeta)e^{-\gamma(\zeta)\hat C}
e^{-\alpha(\zeta)\hat A}e^ {-\beta(\zeta)\hat B},\nonumber
\end{equation}
%
one gets
%
\begin{eqnarray}\label{102}
&&\hspace{-0.8cm}\hat A+\hat B+\hat C-\frac{1}{\delta}\frac{\partial
\delta}{\partial \zeta} = \nonumber\\
&&\hspace{-0.8cm}\frac{\partial \beta}{\partial \tau}\hat
B+\frac{\partial \alpha}{\partial \tau}e^{\beta\hat B}\hat A
e^{-\beta B}+\frac{\partial \gamma}{\partial \tau}e^{\beta\hat
B}e^{\alpha\hat A}\hat C e^{-\alpha\hat A}e^{-\beta\hat B}.
\end{eqnarray}
%
Using the well-known identity for some operators $\hat L_1$ and $\hat L_2$,
%
\begin{equation}\label{103}
e^{t\hat L_1}\hat L_2 e^{-t\hat L_1}= \hat L_2 + t \left[ \hat L_1,\hat L_2 \right] + \frac{t^2}{2} \left[ \hat L_1, \left[ \hat L_1,\hat L_2 \right] \right] + \cdots
\end{equation}
%
we easily obtain
%
\begin{eqnarray}\label{104}
&&e^{\beta\hat B} \hat A e^{-\beta\hat B}=\hat A-\hbar\omega\beta\hat B, \nonumber\\
&& e^{\alpha\hat A} \hat C e^{-\alpha\hat A}= \hat C \exp(-\alpha\hbar\omega).
\end{eqnarray}
%
Consequently,
%
\begin{equation}\label{105}
e^{\beta\hat B}e^{\alpha\hat A}\hat C e^{-\alpha\hat
A}e^{-\beta\hat B}=e^{-\alpha\hbar\omega}\left(\hat C-\beta E^2\right).
\end{equation}
%
%
Substituting Eqs. (\ref{104}) and (\ref{105}) in Eq. (\ref{102}), one gets equations for functions $\alpha(\zeta)$, $\beta(\zeta)$,
$\gamma(\zeta)$ and $\delta(\zeta)$, i.e.
%
\begin{equation}\label{106}
\frac{\partial \alpha}{\partial \tau}=1, \hspace{2mm} \frac{\partial
\beta}{\partial \tau}-\hbar\omega\beta=1, \hspace{2mm} \frac{\partial
\gamma}{\partial \tau}=e^{\hbar\omega\alpha}, \hspace{2mm}
\frac{1}{\delta}\frac{\partial \delta}{\partial
\tau}=E^2\beta. \nonumber
\end{equation}
%
Integrating these equations with the initial conditions
$\kappa(0)=0$, $\beta(0)=0$, $\gamma(0)=0$, $\delta(0)=1$ and substituting the results in Eq.(\ref{100}), one gets the expression
Eq.(\ref{U_II-final}) from the main text.

\section*{Acknowledgements}

This work was supported by the QuantiXLie Centre of Excellence, a project co-financed by the Croatian
Government and European Union through the European Regional Development Fund—the Competitiveness and Cohesion Operational Programme (Grant KK.01.1.1.01.0004), and by IBS-R024-D1.

\end{document}